\documentstyle[11pt,paspconf,psfig]{article}

\begin{document}

\title{Deep multicolour imaging of the field towards the quasar pair PC1643+463~A\&B}

\author{Garret Cotter}

\affil{Royal Greenwich Observatory, 
       Madingley Road,
       Cambridge CB3 0EZ,
       United Kingdom}

\author{Toby Haynes}

\affil{Mullard Radio Astronomy Observatory, 
       Madingley Road,
       Cambridge CB3 0HE,
       United Kingdom}

\begin{abstract}

We present the first results of a deep imaging programme to identify the
system responsible for the Cosmic Microwave Background decrement in the
field towards the $z = 3.8$ quasar pair PC1643+4631~A\&B. Using the prime
focus camera at the William Herschel Telescope we have carried out deep
multicolour optical imaging to search for candidate cluster galaxies at
extremely high redshift. Using $UGR$ colour selection we find the surface
density of $z > 3$ Lyman-break galaxy candidates is at least as great as
that found in the field of the $z = 3.1$ structure discovered by Steidel et
al. (1998), and may be somewhat greater.

\end{abstract}

\section{Introduction}

The field towards the $z = 3.8$ quasar pair PC1643+4631~A\&B contains a
Cosmic Microwave Background (CMB) decrement, believed to be the
Sunyaev-Zel'dovich (SZ) effect of a massive cluster or proto-cluster of
galaxies (Jones et al. 1997; Saunders, these proceedings). Initial optical
and infrared follow-up programme revealed no evidence for cluster galaxies
to at least $z = 1$ (Saunders et al. 1997), and, using an X-ray upper
limit obtained with a deep ROSAT observation, Kneissl et al. (1998) argue
that an isothermal cluster similar those seen at low-redshift clusters
would have to lie at $ z > 2.8 $.  A second candidate high-redshift SZ
cluster, towards a quasar pair at $z = 2.56$, also appears blank in deep
optical and X-ray observations (Richards et al. 1997). We were thus
prompted to initiate a search for cluster galaxies at $ z >> 1$, and as a
first step we have carried out a very deep multicolour optical imaging
programme of the PC1643+4631 field.

\section{Optical imaging}

In 1996 April we observed the central 5' $\times$ 5' region of the
PC1643+4631~A\&B field with the prime focus camera at the 4.2-m William
Herschel Telescope. Images were obtained in $U$,$V$,$R$ and $I$; with a
custom-made $G$ filter kindly lent to us by R.~McMahon; and with two
custom-made narrow-band filters, one centered on Ly-$\alpha$ at the reshift
of the quasars and one centered on the wavelength of the damped Ly-$\alpha$
absorption system at $z = 3.14$ towards quasar PC~1643+463~A. We reached
one-sigma surface brightness limits (in AB mag arcsec$^{-2}$) of $U=29.6$,
$G,R=29.1$, and between 27.4 and 28.6 in the other bands.  Full details of
the observations and data analysis will be published shortly (Haynes et
al. in preparation) and an investigation of the properties of candidate
galaxies in the redshift range $ 1 < z < 3$ is underway (Cotter et al. in
preparation). In this contribution we focus on the candidate members of
what is rapidly becoming a well-studied population: the Lyman-break
galaxies at $z > 3$.

\section{Lyman-break galaxies in the field}

We have followed the principle adopted in the surveys of Steidel et
al. (1995, 1996) to identify galaxies at $z > 3$ via the characteristic
colours caused by the Lyman-limit break straddling the $U$ and $G$
bands. Although we use standard Johnson $U$ and $R$ filters rather than the
custom filters of Steidel et al., we find that the locus of $z > 3$
galaxies is still well defined. As shown in Fig. 1, the  colour criteria $U - G >
2$, $U - G > 4(G - R) + 0.5$ isolate model galaxies at $z > 3$ and avoid
contamination from stars---this region of the colour-colour plane
is equivalent to the ``robust'' criteria of Steidel et al.

\begin{figure}
\centerline{\psfig{file=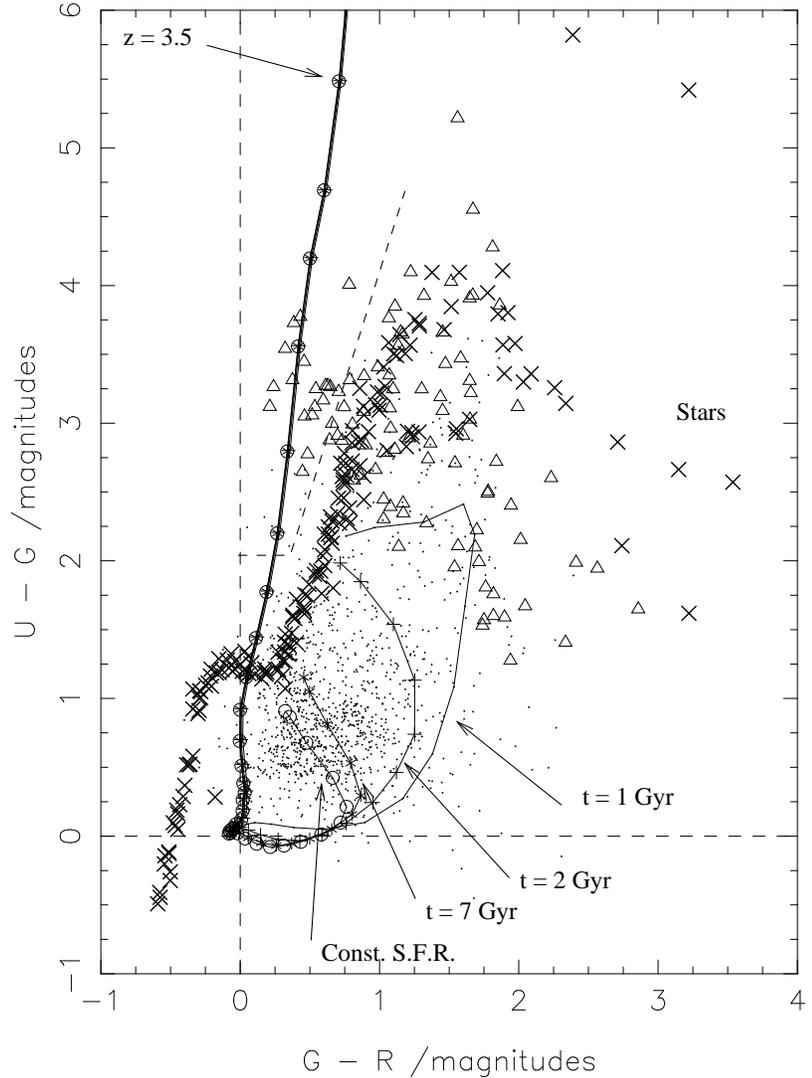,width=0.8\linewidth,clip=,angle=-90}}
\caption{ $UGR$ colour-colour diagram for objects in the central 5'
$\times$ 5' of the PC1643+463 field; compare with, e.g., Fig. 1 of Pettini
et al. (1997). Objects with $R_{\rm AB} < 25.5$ are shown as dots if
detected with one-sigma significance or better in $U$ and as triangles if
undetected to one sigma in $U$. The colours of stars taken from the altas
of Gunn \& Stryker (1983) are shown as crosses. Evolutionary tracks
calculated using the GISSEL96 model spectra of Bruzual and Charlot and
model absorption by intergalactic neutral hydrogen (Madau 1995) are
superimposed. Tracks have been calculated for galaxies forming at redshift
$z = 5$ and having various star-formation histories---constant SFR and
exponentially decreasing SFR with scale times of 1 Gyr, 2 Gyr and 7
Gyr. The model tracks run to $z = 0$ with points plotted at redshift
intervals of 0.1. The dashed line shows the region of colour-colour space
which isolates $z > 3$ model galaxies, equivalent to the ``robust''
selection criteria of Steidel et al.; we find 27 objects in this region.}
\end{figure}

In this region of colour-colour space we find 27 objects (excluding the two
quasars) to $R_{\rm AB} = 25.5$, which gives a surface density of $\approx
1.1 \pm 0.2$ arcmin$^{-2}$. This surface density is comparable to the 0.73
arcmin$^{-2}$ surface density of ``robust'' candidates in the field of the
$z = 3.1$ structure discovered by Steidel et al. (1998); indeed,
tantalizingly, it is greater at the two-sigma level.

Naturally we urge caution in interpreting this result as representing a
real structure at $ z > 3$. Firstly, until the clustering properties of
Lyman-break galaxies are known in detail, it will not be possible to
evaluate the probability that an above-average surface density in a small
field of view corresponds to a real large structure (we note that our
images are several times smaller than those of Steidel et al. 1998). Of
course, prior knowledge of an SZ-producing structure along the line of
sight would increase this probability.  Secondly, an excess of candidate
objects at $z > 3$ does not necessarily imply a structure at this
redshift. Lyman-break galaxies have very small characteristic sizes (see,
e.g., Giavalisco et al. 1996), and so are essentially unresolved with the
typical 0.8'' seeing disk of our images.  Thus, for a fixed surface
brightness limit, one would expect to find an excess of these galaxies if
the field contains an unseen lens at a lower redshift, as in the model
proposed by Saunders et al. (1997).

The only way to determine if these candidates objects represent a real
structure at $z > 3 $ is to proceed with spectroscopy on 10-m class
telescopes. To this end, we are currently planning Keck LRIS observations
in collaboration with the Berkeley group.  Whatever the cause of the excess
of $z > 3$ Lyman-break candidates, it will provide a strong indication of
the nature of the system causing the CMB decrement in the field.

\section*{Acknowledgements}

The investigation of the PC1643+4631 CMB decrement to date has been carried
out by the authors of Jones at al. (1997) and Saunders et al. (1997).

The William Herschel Telescope is operated on the island of La Palma by the
Isaac Newton Group in the Spanish Observatorio del Roque de los Muchachos
of the Instituto de Astrofisica de Canarias.


\begin{references}

\reference Giavalisco, M., Steidel, C.C., Macchetto, F. D. 1996, ApJ, 470, 189

\reference Gunn, J. E., Stryker, L. L. 1983, ApJS, 52, 121 

\reference Jones, M., et al. 1997 ApJ, 479, L1

\reference Kneissl, R., Sunyaev, R. A., White, S. D. M. 1998, MNRAS, submitted

\reference Madau, P. 1995, ApJ, 441, 18

\reference Pettini, M., Steidel., C. C., Adelberger, K., Kellog, M.,
Dickinson, M., Giavalisco, M. 1998, in ORIGINS, ed. J.M. Shull, C.E.
Woodward, and H. Thronson, ASP Conference Series, in press
(astro-ph/9708117)

\reference Richards, E. A.,  Fomalont, E. B., Kellerman, K. I., Partridge, R. B.,
Windhorst, R. A. 1997, AJ, 113, 1475

\reference Saunders, R., et al. 1997 ApJ, 479, L1

\reference Steidel, C. C., Giavalisco, M., Pettini, M., Dickinson, M.,
Adelberger, K. 1996, ApJ, 462, 17L

\reference Steidel, C. C., Pettini, M., Hamilton, D. 1995, AJ, 110, 2519

\reference Steidel, C. C., Adelberger, K., Dickinson, M., Giavalisco, M., 
Pettini, M.,  Kellogg, M. 1998, ApJ, 492, 428

\end{references}
\end{document}